\documentclass[12pt]{article}
\usepackage{geometry}
\usepackage{booktabs}
\usepackage{amsmath}
\usepackage{hyperref}
\usepackage{graphicx}

\usepackage{biblatex}
\usepackage{caption}
\graphicspath{{figures/}}

\title{An improved calendar ring  hole-count\\ for the Antikythera mechanism}
\author{Graham Woan \footnote{ORCID iD 0000-0003-0381-0394}
 \and  Joseph Bayley \footnote{ORCID iD 0000-0003-2306-4106} \\ 
 \small School of Physics and Astronomy, University of Glasgow, Glasgow G12 8QQ, United Kingdom}

\date{2\underline{9} February 2024}

\begin{document}
\maketitle

\abstract{We present a new analysis of the positions of holes beneath the calendar ring of the Antikythera mechanism, as measured by Budiselic et al.\ (2020). We significantly refine their estimate for the number of holes that were present in the full ring. Our $68\%$-credible estimate for this number, taking account of all the data, is $355.24^{ +1.39 }_{ -1.36 }$. If holes adjacent to fractures are removed from the analysis, our estimate becomes $354.08^{ +1.47}_{-1.41}$.   A ring of $360$ holes is strongly disfavoured, and one of 365 holes is not plausible, given our model assumptions.}

\section{Introduction}
The Antikythera mechanism is a multi-component device recovered from a shipwreck close to the Greek island of Antikythera in 1901. It is believed to be the remains of a complex mechanical calculator of ancient origin, and has undergone considerable investigation and analysis to determine its true form and function \cite{2018NatAs...2...35S}.

In a recent paper \cite{budiselic_thoeni_dubno_ramsey_2020}, Budiselic et al.\ presented new, high resolution,  X-ray data on one of the components of the mechanism, the so-called front dial calendar ring, found in Fragment C. Only a part of the full ring survives, and it is fractured into several sections.  Budiselic et al.\ made careful measurements of the positions of closely spaced holes beneath the ring. These holes are thought to have been used to rotationally align the calendar ring, and their number is crucial for the interpretation of the ring's function. The authors generously made their measurements of the hole positions available \cite{DVN/VJGLVS_2019}, and this paper is based entirely on these data.

In this paper, we infer the  number of holes that were present in the complete ring, $N$, given these measurements and some reasonable assumptions.  Budiselic et al.\ presented an analysis that resulted in an estimator for $N$ with a 99\% confidence interval of $346.8$ to $367.2$.  However, using the same data, a clearer and more stringent statement can be made about $N$ using a Bayesian analysis and an improved model for the positional errors in hole placement.  Bayesian methods have several distinct advantages over frequentist methods for addressing problems such as this: first, one can make simple probabilistic statements about the value of $N$ itself, something that frequentist methods are not able to do, by definition.  As a result, there is no need to choose a statistic of the data against which to test a null hypothesis.  It is also straightforward to include all the parameters of the system in the analysis, accounting for the unknown relative orientations of the ring sections.  This allows us to make a concrete statement about $N$ that is constrained by all the pertinent information in the data.

\section{The model}
We assume that originally there were $N$ holes, arranged around a circle of radius $r$.  Today, the circle is partial and fragmented, and exists as a set of $s$ contiguous arc sections that are slightly displaced and rotated with respect to each other.

X-ray images \cite{DVN/VJGLVS_2019}  provide data $\vec{d}_i = (x_i,y_i)$, $1\le i\le n$, on the Cartesian coordinates of $n=81$ contiguous and reportedly  coplanar points  (the hole-centres) that sit on the arcs.    Budiselic et al.\ number these, and we will use their numbering convention for both the holes and sections.  The circle-centres of the section arcs are at $s$ closely bunched but unknown locations $\vec{r}_{0j} = (x_{0j},y_{0j})$, $0\le j\le (s-1)$  and the relative rotations of the sections are also not known precisely.

For the $j$th section, we take $\alpha_j$ as the angular position of the first hole of the full circle, when the section (in its current location and orientation) is extended to that point. Again, the sections only show minimal relative rotation, so we expect these $\alpha_j$ values to be very similar. The apparent angular position of the \(i\)th hole in the $j$th section with respect to its arc-centre is therefore
\begin{equation}
    \phi_{ij} = 2\pi \frac{(i-1)}{N} + \alpha_j.
\end{equation}
There are therefore three unknown parameters for each section, $(x_{0j}, y_{0j},\alpha_j)$, defining a displacement and a rotation in the $(x,y)$ plane. For the moment, we will assume there are no internal distortions of the sections and that they do indeed lie in the $(x,y)$ plane defined by the dataset.

Our goal is to determine the number of holes in the full ring, $N$.  We will use a Bayesian analysis, in which parameters such as $r$ have an associated probability distribution function (PDF) $p(r)$, representing our degree of belief that the parameter lies within any particular range of values, defined as
\begin{equation}
    \text{Prob}(r_1 \le r \le r_2) = \int_{r_1}^{r_2} p(r)\,\mathrm{d}r.
\end{equation}
Our model for the fragments will depend on a multidimensional parameter vector, $\vec{a}$, for which $N$ is just one component. Defining  $ \vec{a} =  (N, r,\{(x_{0j},y_{0j})\},\{\alpha_j\},C) $, the PDF for $N$ alone, given the dataset $ \{\vec{d}_i\}$,  can be computed by marginalising over the other (`nuisance') parameters of the joint PDF, i.e.,
\begin{equation}
    p(N\mid \{\vec{d}_i\}) = \int_{\vec{a}\neg N} p(\vec{a} \mid  \{\vec{d}_i\} )\,\mathrm{d}\vec{a}
    \propto \int_{\vec{a}\neg N} p(\vec{a})\,p(\{\vec{d}_i\}\mid \vec a )\,\mathrm{d}\vec{a},
\end{equation}
where the integrals are  over all the $\vec{a}$ components except $N$, and $C$ represents a parameterised characterisation of random positional errors.  Here, we have used Bayes' theorem,
\begin{equation}
    p(\vec{a} \mid  \{\vec{d}_i\} ) = \frac{p(\vec{a})\,p(\{\vec{d}_i\}\mid \vec a )}{p(\{\vec{d}_i\})},
\end{equation}
recognising that the denominator $p(\{\vec{d}_i\})$ (usually called the `evidence') does not depend on
$\vec{a}$. The $p(\vec{a} \mid  \{\vec{d}_i\} )$ term is usually called the `posterior probability', $p(\vec{a})$ the `prior probability' and $p(\{\vec{d}_i\}\mid \vec a )$ the `likelihood' of the parameters.
Given the relatively tight tolerances in the calendar ring, we will take uniform prior probabilities for $N, r,\{(x_{0j},y_{0j})\}$ and $\{\alpha_j\}$.  We will also initially regard $N$ as a continuous (rather than discrete) parameter, allowing the possibility that the hole spacing had a single discontinuity at a start/end point.

Finally, we will take a Gaussian PDF for the errors in measurement and placement of the holes.
If the intended $i$th hole position in section $j$, relative to its arc centre, is $\vec{r}_{ij}$, we have an error vector,
\begin{equation}
    \vec{e}_{ij} = \vec{r}_{ij} - (\vec{d}_i - \vec{r}_{0j}),
\end{equation}
displacing  the hole from its intended position. It is likely that the ring of holes started as a precisely scribed circle. It is therefore appropriate to differentiate between errors in the radial locations of the holes, and errors around the ring, which we will regard as tangential position errors, and take these as independent. If we define $\hat{\vec{r}}_i$ and $\hat{\vec{t}}_i$ as orthogonal unit vectors aligned with these directions at the intended hole position, the noise covariance matrix in this (locally rotated) coordinate system is simply
\begin{equation}
    C =
    \begin{pmatrix}
        \sigma_\mathrm{r}^2 & 0                   \\
        0                   & \sigma_\mathrm{t}^2
    \end{pmatrix},
\end{equation}
where $\sigma_\mathrm{r}$ and  $\sigma_\mathrm{t}$ are the (unknown) radial and tangential standard deviations.

We can further assume that $\sigma_\mathrm{r,t}$ are the same for every hole and that the hole-to-hole errors are uncorrelated.  In these circumstances the likelihood of the parameters is
\begin{equation}
    p(\{\vec{d}_i\}\mid \vec{a} ) = (2\pi\sigma_\mathrm{r}\sigma_\mathrm{t})^{-n}\prod_{j=0}^{s-1}\prod_{i}^{\text{$i$ in $j$}}
    \exp\left[
        -\frac{(\vec{e}_{ij}\cdot \hat{\vec{r}}_{ij})^2}{2\sigma_\mathrm{r}^2}
        -\frac{(\vec{e}_{ij}\cdot \hat{\vec{t}}_{ij})^2}{2\sigma_\mathrm{t}^2}
        \right],
\end{equation}
where the second product is over the hole indices in the $j$th section, and
\begin{align}
    \hat{\vec{r}}_{ij} & = (\cos\phi_{ij}\,, \sin\phi_{ij}),  \\
    \hat{\vec{t}}_{ij} & = (\sin\phi_{ij}\,, -\cos\phi_{ij}).
\end{align}
As we will see, the errors are tiny in comparison to the radius of the ring, so there is no practical difference between errors referenced to a local tangent and errors along the circular curve.
We will assume non-informative Jeffreys priors on the noise scale parameters $\sigma_\mathrm{r}$ and $\sigma_\mathrm{t}$, (each $\propto 1/\sigma$), though in practice uniform priors on these parameters deliver almost identical results.

\section{Data analysis and results}
The ring fragment is divided into eight sections numbered 0 to 7 (\cite{budiselic_thoeni_dubno_ramsey_2020} Fig.~2), each distinct and with unknown relative translations and rotations. Sections 0 and 4 each contain only one hole.  These sections do not constrain any of our parameters and have therefore been omitted from the analysis.  The remaining six useful sections containing a total of 79 holes, each with $(x,y)$ coordinate data, supplied in millimetres, as constraints.  We have three parameters per section, six usable sections and four further parameters ($N$, $r$, $\sigma_\mathrm{r}$ and $\sigma_\mathrm{t}$), making a total of 22 unknown quantities constrained by the data, and a 22-dimensional posterior space to explore.

We implemented the above analysis using two independently written codes.  The first used the affine-invariant Markov Chain Monte Carlo (MCMC) ensemble sampler \emph{Emcee}~\cite{emcee}, which provides stochastic samples drawn from the joint posterior PDF of the parameters.  These samples provide estimates for  both the full posterior and the parameter marginals, which are the PDFs of parameters individually, taking properly weighted account of the possible values of the other parameters.  The second code used the same model, but explored the posterior space using the \emph{dynesty} nested sampling algorithm~\cite{dynesty} and considered $N$ as a discrete parameter as well as continuous.

We ran Emcee with 100 walkers for $35\,000$ samples, with the first $15\,000$ samples discarded as burn-in.  The final chains were thinned by a factor of 10 before further processing and plotting.  We ran dynesty with 2\,000 live points, stopping when the remaining evidence was less than $1\%$ ($\Delta \ln Z_i < 0.01$) at iteration $i$. We will consider the dynesty results later.

\subsection{Full parameter space results}
Table~\ref{all_results} shows the full set of calendar ring parameter values derived from the MCMC analysis. The results quote the medians of their marginal posteriors, and the $68\%$, $90\%$, $95\%$ and $99\%$ equal-tailed credible intervals.

\begin{table}
    \centering
    \begin{tabular}{cccccc}
        parameter           & median    & $68\%$                   & $90\%$                   & $95\%$                   & $99\%$                   \\ \midrule
        $r$                 & $77.34$   & $^{ +0.29 }_{ -0.28 }$   & $^{ +0.47 }_{ -0.46 }$   & $^{ +0.56 }_{ -0.56 }$   & $^{ +0.75 }_{ -0.74 }$   \\[3pt]
        $N$                 & $355.24$  & $^{ +1.39 }_{ -1.36 }$   & $^{ +2.30 }_{ -2.27 }$   & $^{ +2.75 }_{ -2.73 }$   & $^{ +3.62 }_{ -3.62 }$   \\[3pt]
        $\sigma_\mathrm{r}$ & $0.028$   & $^{ +0.003 }_{ -0.002 }$ & $^{ +0.005 }_{ -0.004 }$ & $^{ +0.006 }_{ -0.004 }$ & $^{ +0.008 }_{ -0.005 }$ \\[3pt]
        $\sigma_\mathrm{t}$ & $0.129$   & $^{ +0.012 }_{ -0.010 }$ & $^{ +0.020 }_{ -0.016 }$ & $^{ +0.025 }_{ -0.019 }$ & $^{ +0.035 }_{ -0.024 }$ \\[7pt]
        $x_{01}$            & $79.69$   & $^{ +0.20 }_{ -0.20 }$   & $^{ +0.33 }_{ -0.33 }$   & $^{ +0.40 }_{ -0.40 }$   & $^{ +0.53 }_{ -0.53 }$   \\[3pt]
        $x_{02}$            & $79.91$   & $^{ +0.23 }_{ -0.22 }$   & $^{ +0.38 }_{ -0.37 }$   & $^{ +0.46 }_{ -0.45 }$   & $^{ +0.61 }_{ -0.59 }$   \\[3pt]
        $x_{03}$            & $79.86$   & $^{ +0.03 }_{ -0.03 }$   & $^{ +0.06 }_{ -0.06 }$   & $^{ +0.07 }_{ -0.07 }$   & $^{ +0.09 }_{ -0.09 }$   \\[3pt]
        $x_{05}$            & $81.44$   & $^{ +1.10 }_{ -1.11 }$   & $^{ +1.84 }_{ -1.84 }$   & $^{ +2.21 }_{ -2.21 }$   & $^{ +2.91 }_{ -2.94 }$   \\[3pt]
        $x_{06}$            & $81.56$   & $^{ +2.46 }_{ -2.41 }$   & $^{ +4.12 }_{ -3.92 }$   & $^{ +4.91 }_{ -4.65 }$   & $^{ +6.51 }_{ -5.96 }$   \\[3pt]
        $x_{07}$            & $83.22$   & $^{ +0.39 }_{ -0.38 }$   & $^{ +0.64 }_{ -0.63 }$   & $^{ +0.76 }_{ -0.75 }$   & $^{ +1.01 }_{ -0.99 }$   \\[7pt]
        $y_{01}$            & $136.03$  & $^{ +0.21 }_{ -0.20 }$   & $^{ +0.35 }_{ -0.34 }$   & $^{ +0.41 }_{ -0.41 }$   & $^{ +0.55 }_{ -0.54 }$   \\[3pt]
        $y_{02}$            & $135.71$  & $^{ +0.27 }_{ -0.27 }$   & $^{ +0.45 }_{ -0.44 }$   & $^{ +0.53 }_{ -0.53 }$   & $^{ +0.70 }_{ -0.71 }$   \\[3pt]
        $y_{03}$            & $135.71$  & $^{ +0.29 }_{ -0.28 }$   & $^{ +0.48 }_{ -0.47 }$   & $^{ +0.57 }_{ -0.56 }$   & $^{ +0.76 }_{ -0.75 }$   \\[3pt]
        $y_{05}$            & $136.10$  & $^{ +0.40 }_{ -0.42 }$   & $^{ +0.66 }_{ -0.69 }$   & $^{ +0.79 }_{ -0.83 }$   & $^{ +1.05 }_{ -1.10 }$   \\[3pt]
        $y_{06}$            & $135.85$  & $^{ +0.80 }_{ -0.86 }$   & $^{ +1.29 }_{ -1.46 }$   & $^{ +1.51 }_{ -1.74 }$   & $^{ +1.95 }_{ -2.30 }$   \\[3pt]
        $y_{07}$            & $136.42$  & $^{ +0.29 }_{ -0.30 }$   & $^{ +0.49 }_{ -0.50 }$   & $^{ +0.59 }_{ -0.59 }$   & $^{ +0.78 }_{ -0.80 }$   \\[7pt]
        $\alpha_1$          & $-145.72$ & $^{ +0.06 }_{ -0.06 }$   & $^{ +0.10 }_{ -0.10 }$   & $^{ +0.12 }_{ -0.12 }$   & $^{ +0.16 }_{ -0.16 }$   \\[3pt]
        $\alpha_2$          & $-145.67$ & $^{ +0.19 }_{ -0.19 }$   & $^{ +0.31 }_{ -0.31 }$   & $^{ +0.37 }_{ -0.37 }$   & $^{ +0.49 }_{ -0.50 }$   \\[3pt]
        $\alpha_3$          & $-145.54$ & $^{ +0.20 }_{ -0.20 }$   & $^{ +0.33 }_{ -0.33 }$   & $^{ +0.39 }_{ -0.39 }$   & $^{ +0.52 }_{ -0.53 }$   \\[3pt]
        $\alpha_5$          & $-146.71$ & $^{ +0.90 }_{ -0.88 }$   & $^{ +1.50 }_{ -1.48 }$   & $^{ +1.81 }_{ -1.77 }$   & $^{ +2.38 }_{ -2.36 }$   \\[3pt]
        $\alpha_6$          & $-146.36$ & $^{ +1.92 }_{ -1.93 }$   & $^{ +3.12 }_{ -3.25 }$   & $^{ +3.71 }_{ -3.86 }$   & $^{ +4.81 }_{ -5.06 }$   \\[3pt]
        $\alpha_7$          & $-147.80$ & $^{ +0.42 }_{ -0.43 }$   & $^{ +0.70 }_{ -0.71 }$   & $^{ +0.83 }_{ -0.85 }$   & $^{ +1.10 }_{ -1.12 }$   \\ \bottomrule
    \end{tabular}

    \caption{The full marginal posterior medians and credible intervals for the inferred parameter values in the ring model. Note that only four parameters, $N$, $r$, $\sigma_\mathrm{r}$  and $\sigma_\mathrm{t}$  are intrinsic to the calendar ring.  The remaining 18 (extrinsic) parameters define the coplanar translational and rotational positions of the six fragments considered.   $N$ is the number of holes in the full ring, $r$ the radius of the full ring, and $\sigma_\mathrm{r}$ and $\sigma_\mathrm{t}$ the standard deviations of the holes from their intended positions in the radial and tangential directions.  $(x_{0j}, y_{0j},\alpha_j)$  are the locations (in millimetres) and rotation angles (in degrees) of the ring sections.}
    \label{all_results}
\end{table}

\begin{figure}
    \centering
    \includegraphics[width=0.5\textwidth]{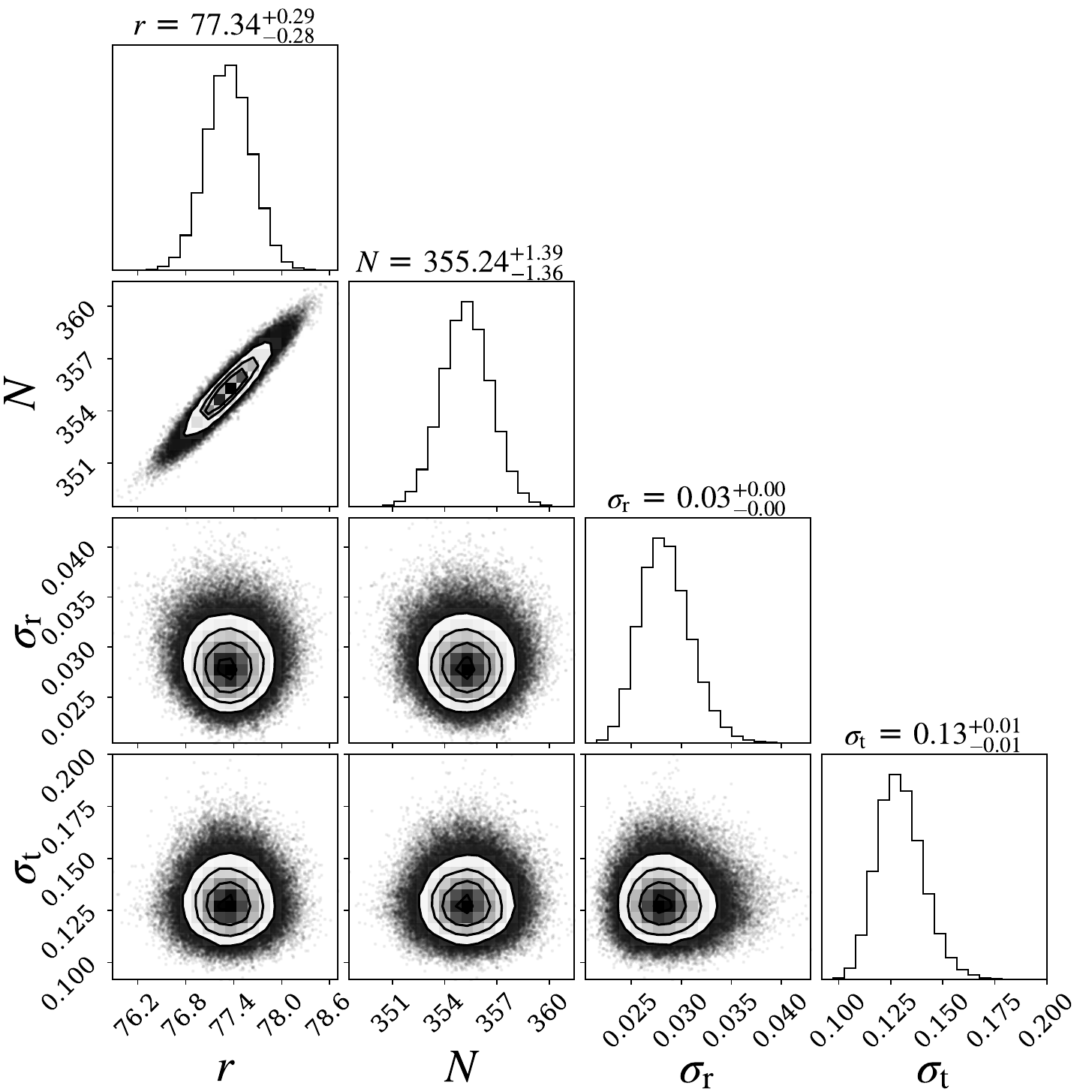}
    \caption{Posterior corner plot for the intrinsic ($N, r, \sigma_\mathrm{r},\sigma_\mathrm{t}$) parameters of the calendar ring, based on all sections except 1 and 4, and marginalised over the other 18 extrinsic parameters.  Length units are millimetres. The expected positive correlation between $r$ and $N$ is clear in their joint marginal plot.}
    \label{small_corner}
\end{figure}

The full posterior sits in a 22-dimensional space, and it is conventional to present it as a plot of one and two-dimensional marginal projections. Even this summary plot is too detailed for display, so in Figure~\ref{small_corner} we show just the plot for the intrinsic parameters $r,N, \sigma_\mathrm{r}$ and $\sigma_\mathrm{t}$.  The nested sampling code delivered results that were fully consistent with these MCMC results.

\begin{figure}
    \centering
    \includegraphics[width=1.0\textwidth]{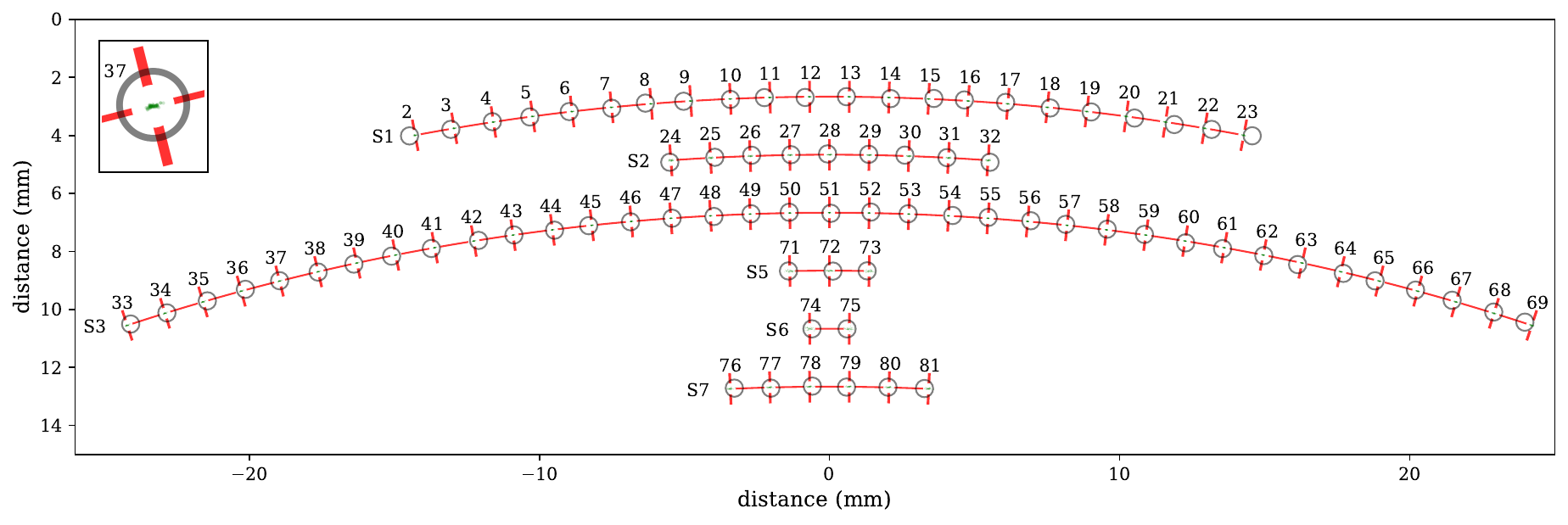}
    \caption{The measured  (grey circles) and  modelled (cross-hairs)  positions of the holes in the calendar ring, broken down by section and using the median marginal values for the model parameters.  The sections are aligned and stacked for ease of viewing.   Fifty randomly drawn posterior predictive values of each hole position are shown in green, with hole 37 magnified to show these values more clearly. Note that $N$ is not constrained to be an integer here.}
    \label{arcs}
\end{figure}

Figure \ref{arcs} shows the hole position data (grey circles) and the median parameter solutions for the hole positions (red cross-hairs and arcs).  The figure also contains 50 posterior predictive values for each hole position.  These positions are computed from representative parameter values drawn from their joint posterior PDF, and will therefore be highly correlated.  The positions of these relative to the actual hole positions highlight the relative magnitude of the radial and tangential errors.  Most of the positional error is tangential, which justifies the use of an anisotropic covariance matrix.

\subsection{Discussion of the full parameter space results}
We performed several additional runs using only subsets of the ring sections.  Unsurprisingly,  most of the information on $N$ and $r$ resides in sections 1, 2 and 3. The other sections are not long enough to constrain $r$ well, and as a result they have relatively little impact on the derived value of $N$. In principle, it is harmless to include these relatively uninformative sections in the analysis, but there is a danger these smaller sections, and the extremities of the larger sections, contain outliers that are not consistent with our assumptions.  We will consider this in Section~\ref{trimmed}.

The results in Table~\ref{all_results} show that the radial error in the hole positions is less than a quarter of the tangential error. It appears that the manufacturer did a better job at putting points on a circle than spacing them evenly, and this insight has a direct bearing on the precision with which we can estimate $r$, and therefore $N$. Our noise model allows for a tighter constraint on $r$ than would be possible with an analysis that does not take this asymmetry into account.
Indeed, an analysis  that does not make this distinction returns an uncertainty in the value of $N$ that is approximately three times higher.  One might imagine that manufacture began with a circle scribed in the metal, using a pair of dividers, and that the holes were marked around this circle with a punch.  If the punch was seated in the scribed groove, the radial error would indeed be very small and largely dependent on the angle at which the punch was struck.  In contrast, each azimuthal position of the punch requires a separate measurement, and would suffer a larger hole-to-hole variation.  However, we note that the degree of manufacturing precision is quite remarkable, with standard errors in hole positions of only 0.028\,mm radially, and 0.129\,mm azimuthally. Budiselic et al.\ quote a standard deviation for their individual position measurements of 0.037\,mm, so a good deal of the radial error may come from the measurements of the X-ray images themselves.

\begin{figure}
    \centering
    \includegraphics[width=0.5\textwidth]{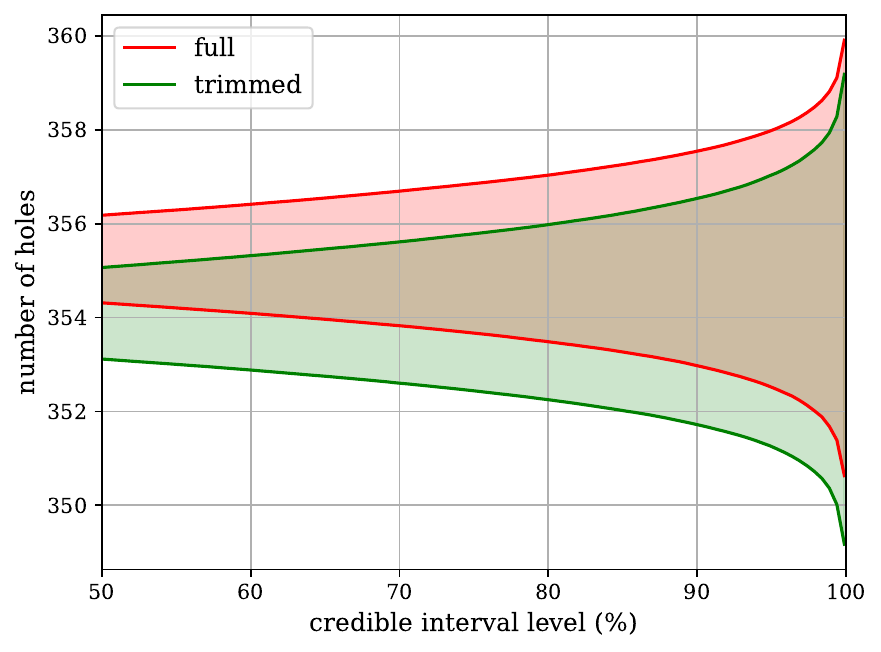}
    \caption{The range of values for the number of holes in the full calendar ring, as a function of the probability that the interval contains the true value.  The result from the full set of holes in section 1, 2, 3, 5, 6 and 7 are in red, and those from the trimmed dataset (section 1, 2, 3 and 7, excluding leading and trailing holes) in green.}
    \label{CI}
\end{figure}

Taking $N$ as a continuous parameter, and with our other assumptions on the noise statistics, our 99\% credible interval for the number of holes in the full circle is $355.24^{ +3.62 }_{ -3.62}$.  The same interval for the radius of the ring is $77.34^{ +0.75 }_{ -0.74 }$\,mm.  This appears to compare very favourably with the results of Budiselic et al., with the proviso that their confidence intervals are not to be interpreted as intervals that contain the true value with the specified confidence.  Only Bayesian statements can be phrased in that way, and, given our assumptions, we are 99\% certain that $N$ lies between 351.62 and 358.86.  The red region in Figure~\ref{CI} shows how the credible interval for $N$ depends on the probability that the truth lies within that interval.

\subsection{Integer values for the number of holes}
\begin{figure}
    \centering
    \includegraphics[width=0.4\textwidth]{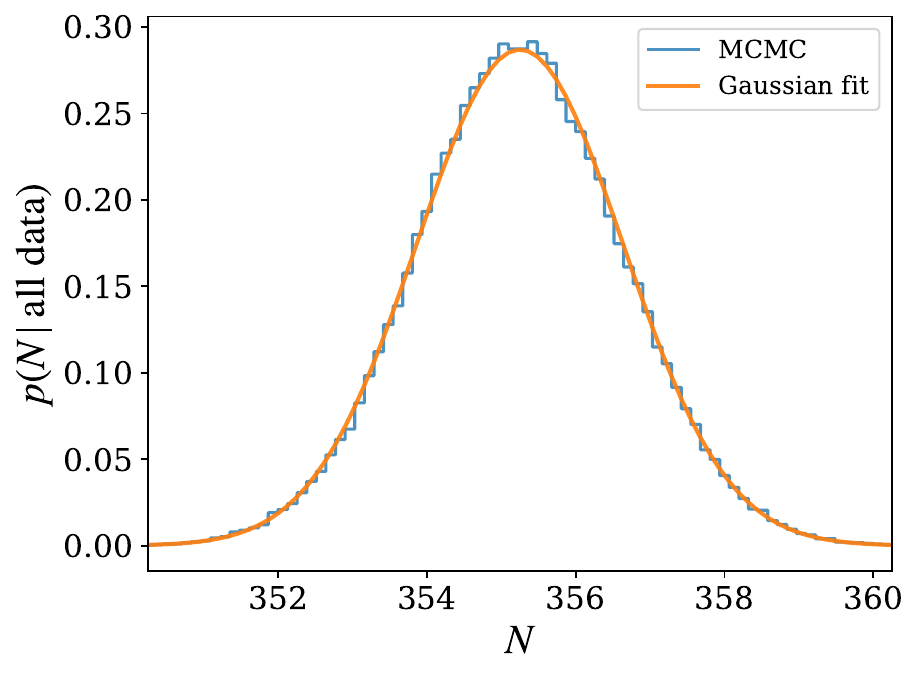}
    \includegraphics[width=0.4\textwidth]{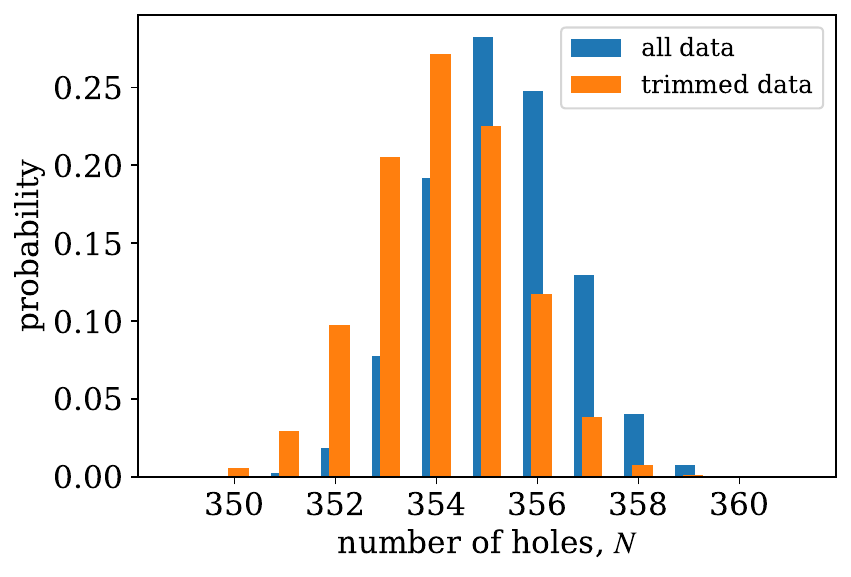}
    \caption{Left: The Gaussian fit to the marginal posterior for $N$, given the full dataset. Right: The data from Table~\ref{probN}, based on Gaussian fits to the full and trimmed datasets, displayed graphically.}
    \label{gaussian_fit}
\end{figure}

Up to this point we have taken $N$ to be a continuous parameter, allowing for a single spacing discontinuity between  holes in a lost section of the ring, or indeed that the original ring was not fully populated with holes.  However, it is reasonable to consider that there was no such discontinuity in the original ring, and that $N$ is an integer. This is equivalent to setting a prior for $N$  that consists of a series of delta functions at discrete integer values. As we initially used a uniform prior for $N$, this new prior can be applied to the posterior post-marginalisation. However, rather few samples will land close to these integer points, so it is useful to take values from a continuous estimate for the probability density. Figure~\ref{gaussian_fit} indicates that the posterior for $N$ is a very close to Gaussian, so  we use a Gaussian as this continuous function,
\begin{equation}
    p(N\mid \{\vec{d}_i\}) =\frac{1}{(2\pi)^{1/2}\sigma_N} \exp\left[ - \frac{(N-\bar N)^2}{2\sigma_N^2} \right],
\end{equation}
characterised by the marginal posterior mean $\bar N= 355.249$, and standard deviation $\sigma_N=1.390$ of the MCMC chain.  We can now compute probabilities for $N$ from this function evaluated at the discrete values singled-out by the prior.  These normalised probabilities, based on all the data, are shown in the centre column of Table~\ref{probN}, and are shown graphically in the blue right-hand plot of Figure~\ref{gaussian_fit}. As a cross-check, we used the nested sampling analysis to compare the relative probabilities of a set of models that assumed integer values for $N$. Again, these produced results that were consistent with Gaussian approximation presented here.

\begin{table}
    \centering
    \begin{tabular}{ccc}
        $N$ & Prob$(N\mid \text{all})$ & Prob$(N\mid\text{trim})$ \\ \midrule
        349 & 0.0000                   & 0.0006                   \\
        350 & 0.0002                   & 0.0055                   \\
        351 & 0.0027                   & 0.0293                   \\
        352 & 0.0187                   & 0.0978                   \\
        353 & 0.0776                   & 0.2056                   \\
        354 & 0.1917                   & 0.2714                   \\
        355 & 0.2823                   & 0.2251                   \\
        356 & 0.2479                   & 0.1173                   \\
        357 & 0.1298                   & 0.0384                   \\
        358 & 0.0405                   & 0.0079                   \\
        359 & 0.0075                   & 0.0010                   \\
        360 & 0.0008                   & 0.0001                   \\
        361 & 0.0001                   & 0.0000                   \\   \bottomrule
    \end{tabular}
    \caption{The probabilities that the full calendar ring contained $N$ equally spaced holes, given the full dataset (column 2), the trimmed dataset (column 3) and our model assumptions.}
    \label{probN}
\end{table}

\subsection{A trimmed dataset}
\label{trimmed}
The preceding analysis has implicitly assumed that the section divisions  identify all the discontinuities in the ring fragment. However, Budiselic et al.\ highlight the difficulty in identifying all of these unambiguously from the X-ray images, and there may be further discontinuities in hole placement that are not listed.   Given the sections are defined by the locations of cracks, we can consider whether the first and last holes of each section are as trustworthy as the rest, as they are by definition adjacent to damage.

Let us therefore consider a reduced, and possibly safer, dataset.  We will exclude the short segments (S5, S6) entirely, and the first and last holes of the remaining sections  S1, S2, S3 and S7 (that is, we exclude holes 2, 23, 24, 32, 33, 69, 76 and 81).   The sections are now shorter and fewer, so we would expect the uncertainties in our derived model parameters to grow slightly.  However, in return, we are somewhat more confident that the data comprises the evenly spaced holes affected by statistically stationary errors assumed by the model.

Table~\ref{reduced} shows the intrinsic parameters derived using this reduced dataset.  The differences are relatively minor, but the values of $r$ and $N$ are slightly reduced.  Although these results are based on fewer holes, they are probably sightly more robust than the full dataset solutions.  The corresponding probabilities of integer $N$ values are shown in the third column of Table~\ref{probN} and the orange bars in the right-hand plot of Figure~\ref{gaussian_fit}.

\begin{table}
    \centering
    \begin{tabular}{cccccc}
        parameter           & median   & $68\%$                   & $90\%$                   & $95\%$                   & $99\%$                   \\ \midrule
        $r$                 & $77.11$  & $^{ +0.30 }_{ -0.29 }$   & $^{ +0.50 }_{ -0.48 }$   & $^{ +0.60 }_{ -0.57 }$   & $^{ +0.81 }_{ -0.77 }$   \\[3pt]
        $N$                 & $354.08$ & $^{ +1.47 }_{ -1.41 }$   & $^{ +2.46 }_{ -2.36 }$   & $^{ +2.96 }_{ -2.83 }$   & $^{ +3.92 }_{ -3.77 }$   \\[3pt]
        $\sigma_\mathrm{r}$ & $0.026$  & $^{ +0.003 }_{ -0.002 }$ & $^{ +0.005 }_{ -0.004 }$ & $^{ +0.006 }_{ -0.004 }$ & $^{ +0.008 }_{ -0.005 }$ \\[3pt]
        $\sigma_\mathrm{t}$ & $0.122$  & $^{ +0.012 }_{ -0.010 }$ & $^{ +0.021 }_{ -0.016 }$ & $^{ +0.025 }_{ -0.019 }$ & $^{ +0.035 }_{ -0.024 }$ \\[3pt]
        \bottomrule
    \end{tabular}
    \caption{The intrinsic parameters of the calendar ring holes, derived from the reduced dataset. Again, lengths are in millimetres.}
    \label{reduced}
\end{table}

\section{Conclusions}
Budiselic et al.\ made careful and precise measurements of hole positions beneath the Antikythera calendar ring, and identified displaced sections of the fragment.  We combined these measurements with a simple model of how they differ from ideal values on a circle.  This was based on a Gaussian distribution that distinguished between the magnitude of error displacements tangential and perpendicular to the circle.  We used this to form a joint likelihood function for the radius of the full circle, the number of holes it contained and the displacement parameters of the broken sections.   After combining this likelihood with uninformative priors, we computed the joint and marginal posterior probability distributions of these parameters using stochastic sampling methods, which we marginalised  to determine the probability distribution for the number of holes.

Budiselic et al.'s paper addressed whether the calendar ring holes represented the 365 days of the Egyptian civil calendar or the 354 days of the lunar calendar. We agree with these authors, that the number of holes beneath the ring is consistent with 354 days, but not with 365.  Additionally, our statement on the number of holes is significantly more constraining.  Using all the data, the 354 hole hypothesis is about 229 times more probable than 360 holes, which they also considered, and vastly more probable than 365 holes.  However, we have not been able to definitively constrain $N$ to a particular integer, or indeed show that it is an integer.

If we remove  holes at the extremities of the sections, that might be affected by fracture, the $68\%$-credible bound on $N$ becomes $354.08^{ +1.47 }_{ -1.41 }$.

A deeper analysis of the dataset is of course possible.  One could relax the coplanar assumption and introduce a small $z$ displacement to each section, and include the remaining two Euler angles for each section's orientation out of the plane.  This would add another 18 parameters to the model.  Additionally, one could use Bayesian methods to infer the number and positions of segment divisions, rather than rely on the choices made by Budiselic et al. These additional degrees of freedom would inevitably increase the uncertainty in the intrinsic parameters.  We note however that the need for these additional degrees of freedom would be revealed by systematic discrepancies between the posterior predictive point-clouds and the measured hole positions, shown in Figure~\ref{arcs}. Given there is no such clear discrepancy, we can assume that the current model has captured the essence of the problem, and that it is unlikely the conclusions would be significantly affected by including them.
\printbibliography
\end{document}